\documentclass[%
 reprint,
superscriptaddress,
 amsmath,amssymb,
 aps,
pra,
floatfix,
]{revtex4-2}

\usepackage{graphicx}
\usepackage{dcolumn}
\usepackage{bm}
\usepackage{hyperref}
\usepackage{braket}
\usepackage{tikz}
\usepackage{quantikz}
\usepackage[subpreambles=false]{standalone}
\usepackage[caption=false]{subfig}
\usepackage{blindtext}
\usepackage{mathdots}

\usepackage{aqarios_colors}

\DeclareMathOperator*{\argmax}{arg\,max}
\DeclareMathOperator*{\argmin}{arg\,min}
        
\begin{document}

\newcommand{\ceillog}[1]{\ensuremath{\lceil\log_2 #1 \rceil}}

\preprint{APS/123-QED}

\title{\textbf{IF-QAOA: A Penalty-Free Approach to Accelerating Constrained Quantum Optimization} 
}%



\author{David Bucher}
\email[Contact To: ]{david.bucher@aqarios.com}
\affiliation{Aqarios GmbH, Munich, Germany}

\author{Jonas Stein}
\affiliation{LMU Munich Department for Computer Science, Munich, Germany}

\author{Sebastian Feld}
\affiliation{Delft University of Technology, Delft, The Netherlands}

\author{Claudia Linnhoff-Popien}
\affiliation{LMU Munich Department for Computer Science, Munich, Germany}

\date{\today}

\begin{abstract}
Traditional methods for handling (inequality) constraints in the Quantum Approximate Optimization Ansatz (QAOA) typically rely on penalty terms and slack variables, which increase problem complexity and expand the search space. More sophisticated mixer-based QAOA variants restrict the search within the feasible assignments but often suffer from prohibitive circuit complexity. This paper presents a low-complexity formalism for incorporating inequality constraints into the cost function of QAOA using an oracle-based subroutine that evaluates constraint satisfaction in an additional register, subsequently called Indicator Function QAOA (IF-QAOA). The IF-QAOA cost function consists of a step-function but does not require a penalty term with additional parameters. Applied to the Knapsack problem, we demonstrate the superior performance of IF-QAOA over conventional penalty-based approaches in simulated experiments. Using advanced QAOA simulation techniques with instances consisting of up to 22 items, we find that IF-QAOA achieves significantly higher solution quality and a faster time-to-solution in 82\% of our benchmark cases. Analysis of the scaling behavior shows favorable scaling of IF-QAOA compared to penalty-based methods. Also, benchmarked against the recently developed Quantum Tree Generator QAOA for Knapsack Problems~\cite{christiansen2024}, we demonstrate higher solution quality for circuits of similar complexity. Additionally, the paper introduces a method for approximate indicator function when the number of ancillary qubits is limited.
With a specialized simulation algorithm based on projective measurements, we empirically demonstrate that a fixed number of ancillary qubits is sufficient to encode general inequality constraints.
\end{abstract}

\maketitle

\section{Introduction}

Recent breakthroughs in quantum computing, from improved coherence times to larger qubit counts, have brought us closer to solving computational problems that remain intractable for classical computers~\cite{acharya2025, liu2025}. Among the most promising applications of Quantum Computing (QC) is their potential to solve hard Combinatorial Optimization Problems (COPs), which lie at the heart of many real-world challenges, ranging from portfolio optimization in finance~\cite{markowitz1952, grant2021, buonaiuto2023} to resource allocation in logistics~\cite{venturelli2016, feld2019a, yarkoni2022a}. While universal fault-tolerant quantum computers remain a future prospect, current Noisy Intermediate-Scale Quantum (NISQ) devices are already being explored for optimization tasks, making efficient quantum optimization algorithms increasingly relevant~\cite{abbas2024}.

COPs represent a fundamental class of computational problems where we seek optimal solutions from a \emph{discrete search space}. 
Formally, (binary) COPs can be defined as finding an optimal solution $x^* \in \{0,1\}^N$ that minimizes an objective function $f(x)$ while satisfying different types of constraints. Those can involve \emph{equality constraints}, meaning a compound quantity $h_j(x)$ has to equal a fixed value, or \emph{inequality constraints}, referring to hard lower (or upper) bounds of $g_i(x)$:
\begin{align}
    x^* = \argmin_x f(x) \quad \text{such that} \quad g_i(x)  &\geq b_i\quad\forall i \nonumber \\ h_j  (x) &= b_j\quad\forall j \nonumber
\end{align}
The objective and constraining functions $f, g_i$ are typically of linear order but can generally take quadratic or even higher-order polynomial form.

Many significant real-world problems alongside Karp's 21 famous NP-complete problems can be formulated with this COP framework~\cite{karp1972, glover2019, lucas2014a}. A particularly important subclass to COPs is the Quadratic Unconstrained Binary Optimization (QUBO), where the objective function is quadratic, and no constraints are present. QUBOs have gained special attention in quantum computing due to their natural mapping to quantum systems through the isomorphism to quadratic Ising Hamiltonians~\cite{lucas2014a}.

This mapping and the computationally challenging task of solving QUBOs, i.e., finding the ground state of Ising Hamiltonians, motivated the development of two major quantum optimization approaches: quantum annealing and gate-based algorithms. Quantum annealers are special-purpose devices that exploit the adiabatic theorem of quantum mechanics~\cite{born1928a, farhi2000} to find the ground state, restricted to solving QUBOs.
In the gate-based paradigm, the primary focus of research is the Quantum Approximate Optimization Ansatz (QAOA), developed by~\citeauthor{farhi2014}~\cite{farhi2014}, which can be seen as a parametrized Trotterization of the annealing process. Parameterization and variational optimization help reduce Trotterization steps, thereby lowering circuit depth, which is inevitable in the NISQ-era of QC. 

Since most industry-relevant COPs feature constraints, they have to be transformed into an unconstrained optimization format to be solved on quantum devices.
While a significant number of problems have been transformed to QUBO~\cite{lucas2014a,glover2019}, this requires reformulating the constraints as \emph{(quadratic) penalty terms} and adding them to the objective function. Despite being generally applicable, this method suffers from increasing the cost landscape's complexity~\cite{shirai2024}. Especially, inequality constraints pose a significant challenge since they also enlarge the search space by introducing slack variables into the optimization problem~\cite{lucas2014a}.

This paper introduces Indicator Function QAOA (IF-QAOA), a novel oracle-based approach for incorporating inequality constraints into QAOA. Our method employs circuit subroutines that encode constraint satisfaction information in an ancillary qubit. This enables the construction of a objective functions where the (negative) cost $f(x)$ is applied conditionally based on constraint satisfaction ($g(x) \geq 0$), namely $\tilde{f}(x) = f(x) \Theta[g(x)]$, where $\Theta$ represents the Heaviside step function.
A key advantage of this formulation is that it eliminates the need for a penalty parameter. 
The IF can be implemented efficiently through Quantum Phase Estimation (QPE) of the constraint function $g(x)$.

While the concept of oracle-based routines in QAOA has already been proposed~\cite{hadfield2017}, previous implementations of those methods for inequality constraints relied on penalty functions~\cite{grandrive2019, kea2023}, using a linear penalty for infeasible assignments. Another implementation~\cite{deller2023} measures the oracle qubit followed by a classically controlled application of one of two cost operators. Consequently, the state collapses to either the feasible or infeasible subspace, potentially hindering optimization.

This work presents a benchmark analysis of IF-QAOA compared to baseline QUBO approaches (slack variables with quadratic penalties) on the Knapsack problem, a COP with a single inequality constraint, addressing instances of up to 22 items through a specialized simulation technique developed for IF-QAOA. While showing significant performance improvements in terms of solution quality and Time-to-Solution (TTS), we also demonstrate improved solution quality compared to the recently developed state-of-the-art Knapsack approach, the Quantum Tree Generator (QTG) QAOA~\cite{christiansen2024}.

Furthermore, we investigate the case where the oracle is approximate, i.e., the QPE cannot resolve the constraining function $g$ due to limited ancillary qubits. A projective measurement helps keep the algorithm well-behaved and allows for efficient simulation of different ancillary register sizes, leading to results that indicate a finite number of qubits can be generally sufficient.

This paper makes the following contributions:
\begin{enumerate}
\item A comprehensive framework for embedding inequality constraints without requiring penalty terms, called IF-QAOA.
\item Efficient simulation techniques for ideal and approximate oracle computations with limited ancillary qubits.
\item Extensive numerical benchmarks on Knapsack problems, demonstrating IF-QAOA's effectiveness compared to conventional quadratic penalty methods recent state-of-the-art QTG-QAOA.
\end{enumerate}

The paper is organized in the following way: Sec.~\ref{sec:background} introduces relevant background on quantum optimization, and Sec.~\ref{sec:related-work} reviews previous approaches to constraint handling in quantum optimization. Sec.~\ref{sec:methods} presents our indicator function methodology and theoretical framework. Sec.~\ref{sec:numerical} provides the numerical experiments and analyzes the performance of our approach in both exact and approximate implementations. Finally, Sec.~\ref{sec:discussion} concludes with a discussion of practical implications and future directions.



\section{Background}\label{sec:background}

\subsection{Quantum Approximate Optimization Algorithm}

\citeauthor{farhi2014} first introduced the QAOA to find the maximum cut in three-regular graphs~\cite{farhi2014}. Nevertheless, QAOA can be applied to find the optimal solution to any QUBO or even higher-order optimization problem~\cite{glos2022}. The algorithm is initialized in an equal superposition of $N$ qubits $\ket{+}^{\otimes N}$ for a problem consisting of $N$ binary decision variables. This is followed by $p$ repetitions of two unitary operations in alternating order, designed to progressively the probability of measuring an optimal bit-string $x^*$ upon final measurement: 
\begin{align}
    \ket{\bm{\beta}, \bm{\gamma}} = U_M(\beta_p) U_f(\gamma_p) \cdots U_M(\beta_1) U_f(\gamma_1) \ket{+}^{\otimes N},
\end{align}
where every layer is parameterized by the angles $\beta_i$ and $\gamma_i$.  The cost layer is given by the exponentiation of the diagonal cost Hamiltonian $H_f = \sum_x f(x)\ket{x}\bra{x}$
\begin{align}
    U_f(\gamma) &= \exp(-i \gamma H_f).
\end{align}

The mixer unitary $U_M$ is the exponentiation of the standard mixer Hamiltonian $H_M = \sum_i \sigma_x^{(i)}$:
\begin{align}
    U_M(\beta) &= \exp(-i \beta H_M) = \prod_k R_X^{(k)}(2\beta),
\end{align}
where $R_X^{(k)}$ is the Pauli-$X$ rotation on qubit $k$.

The QAOA circuit consists of the free rotation angles $\bm{\beta}$ and $\bm{\gamma}$, which must be explicitly selected for each problem instance to ensure effective optimization. This property classifies QAOA as a variationl quantum algorithm. We aim to find parameters that lead to a high probability $P^* = \left| \braket{x^* | \bm{\beta}, \bm{\gamma}} \right|^2$ of measuring an optimal bit string $x^*$ at the end of the algorithm. For cases where $H_f$ has a degenerate ground state, we aggregate the probabilities of measuring any ground state into $P^*$. However, since $x^*$ is not accessible a priori, optimizing for $P^*$ is infeasible. Consequently, QAOA minimizes the expectation value of the cost Hamiltonian $H_f$ instead: 
\begin{align}
    C(\bm{\beta}, \bm{\gamma}) = \bra{\bm{\beta}, \bm{\gamma}} H_f \ket{\bm{\beta}, \bm{\gamma}} = \sum_{x = 0}^{2^N -1} f(x) \left| \braket{x | \bm{\beta}, \bm{\gamma}} \right|^2.
\end{align}
The variational protocol embeds the quantum algorithm within a classical loop for optimizing the continuous parameters
\begin{align}
    \min_{\bm{\beta}, \bm{\gamma}} C(\bm{\beta}, \bm{\gamma}).
\end{align}
When the type of classical optimizer chosen requires gradients to be computed, the parameter shift rule or adjoint differentiation are typically employed~\cite{crooks2019, jones2020}.

Since QAOA can be seen as a Trotterization of the adiabatic algorithm for optimization problems, we can also predefine the angles to growing $\gamma_i$'s and shrinking $\beta_i$'s. This initialization strategy has been successfully demonstrated for QAOA, even for a relatively low number of QAOA layers $p$~\cite{sack2021, montanez-barrera2024}.

\subsection{Knapsack Problem}
The 0-1 Knapsack problem is the most basic example of a combinatorial optimization problem with a single inequality constraint. The problem is defined as finding the optimal selection of items $I^* \subseteq I, |I| = N$ that maximizes its total \emph{value} while adhering to an absolute \emph{weight capacity} $W > 0$. Each item $(w_i, v_i) \in I$ is characterized by its individual weight $w_i \geq 0$ and a value $v_i \geq 0$. Generally, $w_i$, $v_i$ and $W$ are considered integer-valued. Formalized, this optimization can be expressed as follows:
\begin{align}
    \argmax_{x \in \{0, 1\}^N} \sum_i v_i x_i \quad \text{such that} \quad \sum_i w_i x_i \leq W,
    \label{eq:knapsack-problem}
\end{align}
where the set of binary variables $x_i$ indicates whether the $i$-th item is selected ($x_i = 1$) or not ($x_i= 0$)~\cite{kellerer2004}.

The Knapsack problem belongs in the NP-hard complexity class, implying that it is at least as computationally difficult as the hardest problem within NP~\cite{garey1990}. As a consequence, assuming P $\neq$ NP, no polynomial-time algorithm can exist to solve the Knapsack problem optimally for all instances. Yet, in practice, many practically relevant Knapsack instances can often be solved efficiently within polynomial time scaling using dynamic programming or branch-and-bound methods~\cite{martello1999, kolesar1967}.
The problem finds numerous applications, for instance, in financial portfolio optimization~\cite{kellerer2004}, and serves as a foundation for various logistical and operational optimization problems, including the bin-packing problem~\cite{martello1990}.

We can generalize Eq.~\eqref{eq:knapsack-problem} into the following in form:
\begin{align}
    \argmin_{x\in\{0,1\}^N} f(x) \quad \text{s.t.} \quad g_i(x) \geq 0 \quad \forall i \in 1,\dots,k,
    \label{eq:general-optproblem}
\end{align}
which allows us to consider any inequality-constrained binary programming problems. In the Knapsack case, $k = 1$, $f(x) = -v^T x$ and $g_1(x) = W - w^Tx$. We will subsequently call the $\mathcal{D} = \{0, 1\}^N$ the \emph{domain space} of the problem and $\mathcal{D} \supseteq \mathcal{F} = \{x \in \mathcal{D}\,|\, g_j(x) \geq 0\, \forall j\}$ the \emph{feasible subspace}. Let us also assume w.l.o.g. that $f(x) \leq 0 \,\forall x \in \mathcal{D}$, which is naturally the case with Knapsack and can generally be ensured by subtracting an upper bound of $f$ from itself.

\section{Literature Review}
\label{sec:related-work}

Before we delve into an in-depth review of constraint satisfaction methods for QAOA, we briefly discuss alternative quantum optimization algorithms that can include constraints in the optimization. Based on one of the most prominent algorithms for theoretical quantum advantage, the Grover algorithm for unstructured search~\cite{grover1996}, the quantum minimum finding has emerged as a modification for solving general optimization problems~\cite{durr1999}. The oracle-based algorithm was later extended also to feature general constraints~\cite{gilliam2021}. Recently, quantum minimum finding has been successfully employed in combination with Quantum Tree Generators (QTGs) to solve the Knapsack problem specifically~\cite{wilkening2024}.
Other ansätze feature hybrid methods that ensure constraint satisfaction through classical post-processing~\cite{shirai2024a} or an iterative quantum-classical hybrid
loop~\cite{gambella2020a, zhao2022}.

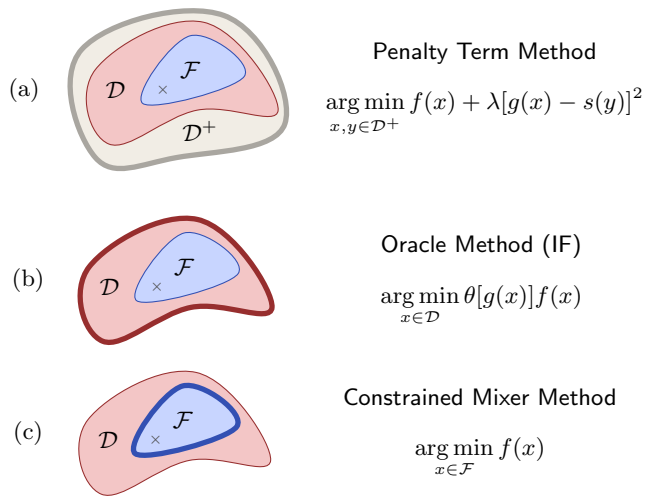
\begin{figure}
    (a)
    \begin{minipage}[c][3cm]{3.5cm}
    \centering
    \begin{tikzpicture}[scale=0.9]
\def\numPoints{20}
\def\radius{2}

\filldraw[fill=sand, draw=sand!70!black, line width=2pt] plot[smooth cycle, tension=0.7] coordinates {  (-1.2, -1.2) (-1.4, 0.4) (0.7, 0.9) (1.6, -0.7) (0, -1.2) };

\filldraw[fill=rocketfire!30, draw=rocketfire!70!black] plot[smooth cycle, tension=0.7] coordinates { (-1, -1) (-1.2, 0.1) (0.4, 0.8) (1.5, -0.5) (0.3, -0.5) };

\filldraw[fill=aquarius!30, draw=aquarius!70!black] plot[smooth cycle, tension=0.7] coordinates { (-0.5, -0.4) (0.2, 0.6) (1, 0) };

\node[black!70] at (-0.2, -0.2) {\tiny $\times$};

\node at (0.2, 0.1) {$\mathcal{F}$};
\node at (-0.9, -0.2) {$\mathcal{D}$};
\node at (0.35, -0.8) {$\mathcal{D}^{+}$};

\end{tikzpicture} 
    \end{minipage}
    \begin{minipage}[c][3cm]{4.4cm}
    \textsf{Penalty Term Method}
    \[\argmin_{x,y \in \mathcal{D}^+} f(x) + \lambda [g(x) - s(y)]^2 \]
    \end{minipage}\\
    (b)\begin{minipage}[c][2cm]{3.5cm}
    \centering
    \begin{tikzpicture}[scale=0.9]
\def\numPoints{20}
\def\radius{2}

\filldraw[fill=rocketfire!30, draw=rocketfire!70!black, line width=2pt] plot[smooth cycle, tension=0.7] coordinates { (-1, -1) (-1.2, 0.1) (0.4, 0.8) (1.5, -0.5) (0.3, -0.5) };

\filldraw[fill=aquarius!30, draw=aquarius!70!black] plot[smooth cycle, tension=0.7] coordinates { (-0.5, -0.4) (0.2, 0.6) (1, 0) };

\node[black!70] at (-0.2, -0.2) {\tiny $\times$};

\node at (0.2, 0.1) {$\mathcal{F}$};
\node at (-0.9, -0.2) {$\mathcal{D}$};

\end{tikzpicture}
    \end{minipage}
    \begin{minipage}[c][2cm]{4.4cm}
    \textsf{Oracle Method (IF)}
    \[\argmin_{x \in \mathcal{D}} \theta[g(x)] f(x) \]
    \end{minipage}\\
    (c)\begin{minipage}[c][2cm]{3.5cm}
    \centering
    \begin{tikzpicture}[scale=0.9]
\def\numPoints{20}
\def\radius{2}

\filldraw[fill=rocketfire!30, draw=rocketfire!70!black] plot[smooth cycle, tension=0.7] coordinates { (-1, -1) (-1.2, 0.1) (0.4, 0.8) (1.5, -0.5) (0.3, -0.5) };

\filldraw[fill=aquarius!30, draw=aquarius!70!black, line width=2pt] plot[smooth cycle, tension=0.7] coordinates { (-0.5, -0.4) (0.2, 0.6) (1, 0) };

\node[black!70] at (-0.2, -0.2) {\tiny $\times$};

\node at (0.2, 0.1) {$\mathcal{F}$};
\node at (-0.9, -0.2) {$\mathcal{D}$};

\end{tikzpicture} 
    \end{minipage}
    \begin{minipage}[c][2cm]{4.4cm}
    \textsf{Constrained Mixer Method}
    \[\argmin_{x \in \mathcal{F}} f(x) \]
    \end{minipage}\\
    \caption{Showcasing the search spaces of the various optimization strategies. In (a), the search space is enlarged to $\mathcal{D}^+$ through slack variables in the penalty term. In (b), the search space is exactly the space of the originally defined problem, and in (c), the search space is effectively reduced to the feasible subspace. The shapes of the domains are exemplary.}
    \label{fig:search-space}
\end{figure}


Now, we systematically review the various methodologies developed for handling constraints in QAOA, focusing mainly on inequality constraints that arise naturally in problems such as Knapsack.

\subsection{Penalty-Based Methods}
The conventional approach to incorporating constraints into QAOA involves reformulating them into QUBO format and augmenting the objective function with a quadratic penalty~\cite{lucas2014a}. 

Taking Knapsack as an example, this reformulation requires introducing ancillary binary variables (slack variables) to encode the feasible region of the inequality constraint. Specifically, the interval $[0, W]$ must be represented through slack variable assignments. The logarithmic integer encoding~\cite{lucas2014a} provides a qubit-efficient representation through a slack term $s(y) = \sum_j a_j y_j$,
where $a_j = 2^j$ for $j < M$, and $a_M = W - 2^{M-1} + 1$ with $M =
\ceillog{W}$. The resulting QUBO formulation becomes:
\begin{align}\label{eq:slack-penalty}
    \min_{x, y \in \mathcal{D}^+} f(x) + \lambda [g(x) - s(y)]^2,
\end{align}
where $\lambda > 0$ is a penalty coefficient.

This penalty-based approach presents two significant challenges. First, it introduces an additional hyperparameter $\lambda$ that needs to ensure no infeasible state has lower energy than a feasible one. With the square of the constraint violation, the penalty term can overshadow the initial problem and significantly skew the energy spectrum, leading to a more complex energy landscape to optimize~\cite{mirkarimi2024}. Second, slack variables expand the search space from $\mathcal{D}$ to $\mathcal{D}^+ = \mathcal{D} \times \{0, 1\}^M$, as illustrated in Fig.~\ref{fig:search-space}a, making the QUBO considerably harder to solve.

While these drawbacks are inherent to QUBO-based optimization methods, several techniques have been proposed to mitigate them.
Cross-entropy methods can help determine appropriate penalty factors~\cite{roch2021}, although at the cost of increased computational overhead. Recent numerical studies suggest that incorporating the penalty term in the quantum routine while excluding slack variable assignments from the classical expectation value calculation yields promising results~\cite{hess2024}.

Alternatively, slack-variable-free approaches have emerged in literature: \citeauthor{gabbassov2024}.~\cite{gabbassov2024} propose incorporating the Knapsack constraint through Lagrangian duality, eliminating the need for slack variables. While Lagrangian methods typically require iterative penalty updates, the authors argue that in approximate optimization scenarios, a well-chosen initial penalty coefficient suffices for boosting the probability of sampling optimal or near-optimal solutions. Similarly, \citeauthor{montanez-barrera2024a}~\cite{montanez-barrera2024a} propose unbalanced penalization that approximates an exponential barrier using a linear and quadratic term without slack variables. Van Dam et al.~\cite{dam2022} present another slack-free approach, specifically for Knapsack, that warm-starts QAOA with a greedy initial solution and uses specialized copula mixer Hamiltonians to bias the exploration towards the distribution generated by the greedy solution. Numerically, all three approaches manage to demonstrate significant improvements over baseline methods.

\subsection{Oracle Methods}
Generally, objective functions in QAOA are of QUBO or sometimes higher-order polynomial form~\cite{glos2022}. However, it is possible to use the full capabilities of gate-based QC to construct more complex cost functions. \citeauthor{hadfield2017}~\cite{hadfield2017} proposed using an oracle subroutine: a QPE as part of the cost unitary routine that synthesizes information about constraint satisfaction onto the state of a single qubit, which can be seen as an \emph{indicator} qubit. While concrete constructions of this method are not further followed in~\cite{hadfield2017}, Refs.~\cite{grandrive2019, kea2023} explore this path to solve the Knapsack problem. Ref.~\cite{grandrive2019} utilizes arithmetic adder circuits to compute $g(x)$ on separate registers, and Ref.~\cite{kea2023} uses the QPE/QFT-adder based technique~\cite{draper2000}. Both methods implement the following unconstrained non-linear cost function
\begin{align}
    f(x) + \lambda (1 - \Theta[g(x)]) g(x),
\end{align}
adding a linear penalty if the constraint is violated, where $\Theta$ is the Heaviside step function, here defined to be $\Theta[g(x)] = 1$ if $g(x) \geq 0$ and zero otherwise. The method investigated by Ref.~\cite{kea2023} is similar to IF-QAOA, except our method does not require a penalty setting, as explained in Sec.~\ref{sec:methods}. While Refs.~\cite{grandrive2019,kea2023} demonstrate small-scale numerical success of their method, they lack an extensive numerical analysis.

In principle, oracle-techniques search the entire domain space of the problem $\mathcal{D}$. They can, therefore, be considered more efficient than the slack-variable methods, see Fig.~\ref{fig:search-space}b.
A similar QPE-based oracle concept has been brought forward by Ref.~\cite{deller2023} in the context of constrained qudit (quantum digit) optimization. However, the authors propose a measurement of the indicator qubit, leading to the state collapsing to either the feasible or infeasible subspace of the Hilbert space. A cost Hamiltonian with penalization is applied if the indicator qubit measurement reveals the infeasible region. Then, the statevector is in a superposition of \emph{only} infeasible states, which the subsequent QAOA iterations must revert. This raises the question of whether restarting upon failed measurement is more efficient. IF-QAOA is entirely cost-function-based and does not rely on measurements (in the ideal case).

\subsection{Constraint Preserving Mixers:}

Constraint-preserving techniques represent a class of QAOA circuit architectures
that operate entirely within the feasible Hilbert subspace
\begin{align}
    \mathcal{H}_\mathcal{F} = \{\ket{\psi} \in \mathcal{H} 
    \,|\,\braket{x|\psi} = 0\,\forall x \in \mathcal{D} \setminus \mathcal{F}\}.
\end{align}

While these methods explore the smallest possible search space by construction
(see Fig.~\ref{fig:search-space}), the required circuit complexity to enforce
constraints can be substantial. This creates a fundamental trade-off between
search space size and implementation cost that varies significantly with the
type of constraint being enforced.

\paragraph{Quantum Alternating Operator Ansatz:}
\citeauthor{hadfield2019a}~\cite{hadfield2019a} derived that QAOA mixers need not be Hamiltonian evolutions but can be arbitrary unitaries that enable transitions between all feasible states. This insight allows for the design of constraint-preserving mixers that prevent transitions from feasible to infeasible states, formally expressed as $\bra{x}U_M(\beta)\ket{y} = 0, \forall \beta, x\in \mathcal{F},y\notin\mathcal{F}$. The $XY$-mixer exemplifies this approach, effectively enforcing one-hot constraints (Hamming weight 1 bit-strings)~\cite{wang2020}.

Hamming weight constraints, i.e., max-$k$-vertex covering constraints~\cite{cook2020}, share a vital property, making the construction of mixer routines feasible. Namely, their solution structure is known beforehand. In contrast, inequality constraints lack inherent structure and depend heavily on constraint-specific data, leading to substantially more challenging construction of mixer operators. While methods exist for inequality constraints, they often incur prohibitive implementation costs, especially for non-sparse cases~\cite{sawaya2023a, fuchs2022, ruan2020}.

Ref.~\cite{shirai2024} introduced a constructive approach that maps the feasible subspace onto a reduced set of qubits with matching Hilbert space dimension. Each mixer operation combines the mapping unitary, standard $X$-mixer on the reduced qubits, and the inverse mapping. While effective for structured constraints, inequality constraints require a machine learning method to determine the mapping unitary, which the authors also successfully demonstrated. However, the variational ansatz is never guaranteed to produce an utterly valid mapping unitary. Furthermore, this method introduces an additional variational optimization loop before QAOA optimization, increasing runtime overhead.

\paragraph{Grover Mixers:}
Based on the works of Ref.~\cite{hadfield2019a}, \citeauthor{bartschi2020}~\cite{bartschi2020} developed a novel mixer method using a circuit for preparing the equal superposition of all feasible states. By combining this state preparation circuit with the Grover diffusion operator, they construct the following mixer
\begin{align}
    U_M(\beta) = \exp\left(-\frac{i \beta }{|\mathcal{F}|}\sum_{x,y\in\mathcal{F}} \ket{x}\bra{y}\right),
\end{align}
that restricts probability transfer between feasible states only. This ansatz effectively shifts the complexity of synthesizing an appropriate mixer operator to the state preparation circuit. For constraints yielding regular state patterns, such as one-hot constraints~\cite{cruz2019}, efficient implementations exist through short-depth state preparation circuits. However, generating an equal superposition of solely feasible solutions under an inequality state may be computationally hard~\cite{sawaya2023a} (e.g., the number of feasible solutions has to be known beforehand).

QTG (Quantum Tree Generator) offers a solution by constructing non-uniform superpositions of feasible states for single inequality constraints, namely the Knapsack problem~\cite{wilkening2024}. While initially intended to be used in the Grover adaptive search context, the superposition state can also be used to construct a Grover-like mixer for the QAOA, as demonstrated successfully in Ref.~\cite{christiansen2024}. The authors coined the term AAM-QAOA because of the non-uniform superposition and proved the convergence of such mixers. Numerical experiments have shown significant performance improvements compared to the results from Ref.~\cite{dam2022}.

\paragraph{Zeno Dynamics:}
\citeauthor{herman2023}~\cite{herman2023} proposed a constraint mixer for inequality constraints using the default $X$-mixer and quantum Zeno dynamics, where repeated projective measurements constrain the quantum state within the feasible subspace. While this method has been successfully demonstrated numerically and on trapped-ion hardware, solving the portfolio optimization problem, the Zeno approach requires dissecting the mixer operator into small time steps. In that way, Zeno dynamics increases the probability of successful projective measurements on indicator qubit. The measurement circuit is similar to the one explored in Ref.~\cite{deller2023}. This dissection leads to lengthy circuits that may be obstrucive in delivering a general performance advantage. There is a trade-off between circuit depth and overall success probability.

\section{Methods}\label{sec:methods}

\subsection{Single Constraint Algorithm}\label{sec:ifalgo}

\begin{figure*}
    \subfloat[IF-QAOA circuit\label{fig:ifcirc}]{\includestandalone{drawings/qaoa_qpe}}
    \subfloat[IF phase factor\label{fig:phasefactor}]{\includegraphics[scale=0.8]{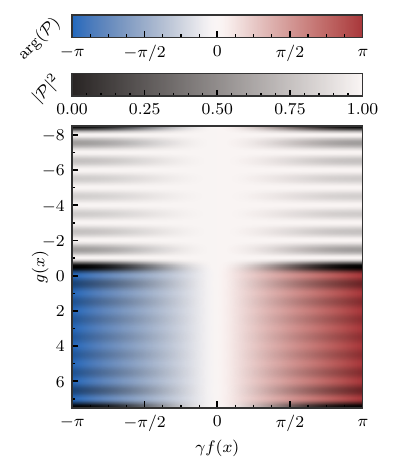}}
    
    \caption{Panel (a) shows the IF-QAOA circuit. The cost layer consists of the QPE, evaluating whether constraint $g$ has been satisfied, and a controlled application of the cost Hamiltonian based on the indicator qubit. The projective measurement on the QPE register is only required when $g$  is not resolvable by the QPE register. Panel (b) presents the projection factor $\mathcal{P}$, defined in Eq.~\eqref{eq:projector-defined}, in the approximate application of the indicator function, when $g$ is not resolvable by the QPE-register, here for $M=4$. The shading of the plot shows the probability of successful projective measurements. Clearly, when $g$ is an integer, this probability is always one. The hue indicates the applied phase to the compute register, showing that the application only happens for $g(x) \geq 0$.}
    \label{fig:methods}
\end{figure*}

The IF-QAOA is a constraint integration method that only applies the (polynomial) cost function $f(x) \leq 0 \ \forall  x \in \{0, 1\}^N$ to the feasible solution candidates without including slack variables in the optimization problem but using ancillary qubits for oracle construction. Let us consider an optimization problem with a single inequality constraint $g(x) \geq 0$ with an integer-valued image $g(\mathcal{D}) \in \{g^-,\dots,g^+\} \subset \mathbb{Z}$, containing defined upper and lower bounds. In the case of linear $f$ and $g$, this case resembles the Knapsack problem with integer weights. We define the minimization objective for the IF-QAOA as follows
\begin{align}\label{eq:if-cost}
    \tilde{f}(x) = f(x) \Theta[g(x)].
\end{align}
Since $\Theta[g(x)] = 1$ if $g(x) \geq 0$ and, therefore, $x \notin \mathcal{F}$, $\tilde{f}(x) \leq \tilde{f}(y) =0\,\forall x\in\mathcal{F}, y \notin \mathcal{F}$, because $f(x) \leq 0\, \forall x \in \mathcal{D}$.

Therefore, the cost layer routine of the IF-QAOA circuit needs to have the following effect on the phase
\begin{align}\label{eq:ifqaoa-phase}
    U_{\text{IF-}f}(\gamma) \ket{x} = e^{-i\gamma \tilde{f}(x)}\ket{x}.
\end{align}

To assemble the phase circuit, we construct the constraint Hamiltonian $H_g = \sum_x g(x)\ket{x}\bra{x}$, diagonal in the computational basis, which can subsequently be used to perform QPE with $M$ qubits in an ancillary register on the unitary $U_g = e^{2 \pi i H_g / 2^M}$. Gate level implementation of the $U_g$ is possible through multi-controlled phase gates, as long as $g$ is polynomial. The number of ancillary qubits $M$ is defined by the resolution required for $g$. Since $g(x) \in \mathbb{Z}$, we define $M = \max\{\ceillog {|g^-|}, \ceillog{(g^+ + 1)}\} + 1$ to fully resolve both the positive and negative part of $g$. Finally, we continue applying the QPE
\begin{align}
& \mathrm{QPE}_{U_g}\ket{x}\ket{0}= \mathrm{QFT}^\dagger\frac{1}{\sqrt{2^M}}\sum_k e^{\frac{2\pi i}{2^M} g(x) y }\ket{x}\ket{k} 
\\ &= \frac{1}{2^M} \sum_{k} \sum_{z} e^{-\frac{2 \pi i}{2^M} (z - g(x)) y } \ket{x}\ket{z} \label{eq:ex-qft-applied}
\\ &= \sum_{z} \delta_{g(x), z} \ket{x} \ket{z} =  \ket{x}\ket{g(x)} \label{eq:ex-delta}.
\end{align}
Since the output register of the QPE is in two's complement~\cite{nielsen2012}, we define $\ket{g(x)} = \ket{z_1 \cdots z_M}$, with $g(x) = 2^M (-z_1/2 + \sum_{i=2} 2^{-i}z_i)$. Consequently, the sign of $g(x)$ is indicated by the first qubit $z_1$, i.e., $g(x) \geq 0 \Leftrightarrow z_1 =0$.

Based on that state, we can apply the ordinary cost unitary inversely controlled by the indicator qubit $z_1$ and afterwards uncompute the ancillary QPE register to be reused in the following QAOA iterations:
\begin{align}
    U_{\text{IF-}f}(\gamma) = \mathrm{QPE}_{U_g}^\dagger \bar{C}_{z_1}(U_f(\gamma))\mathrm{QPE}_{U_g},
\end{align}
where $\bar{C}_{z_1}$ denotes the negated control on $z_1$, as visualized in the circuit Fig.~\ref{fig:ifcirc}. Note that the circuit in Fig.~\ref{fig:ifcirc} additionally features a projective measurement of the QPE register, which is needed when $g(x)$ will not be resolvable on the QPE register, e.g., when the image of $g$ is non-integer, as detailed in Sec.~\ref{sec:approximate}. In the resolvable case discussed in this section, the measurements will always be zero and can, therefore, be omitted.

\subsubsection*{Resource Estimation}
Let us compare the resources required to implement IF-QAOA against the quadratic penalty QUBO by assuming $f$ and $g$ are linear binary functions. The number of ancillary qubits required differs: generally, $M_\text{IF} \geq M_\text{QUBO}$, since the IF-QAOA QPE register needs to resolve the whole cost function. In contrast, the QUBO method only requires encoding the feasible part of $g$. Please refer to App.~\ref{sec:app-clops} for detailed resource estimations. The resources to implement the QUBO penalty approach are: $(N+M_\text{QUBO})$ layers with $(N+M_\text{QUBO}) (N+M_\text{QUBO} - 1) / 2$ controlled phase rotations plus a single layer of phase gates.
The IF-QAOA consists of the QPE part, needing $N$ layers and $N M_\text{IF}$ controlled phase gates for the phase addition, followed by $2M_\text{IF}-1$ QFT layers with $M_\text{IF} (M_\text{IF} + 1)/2$ gates. The controlled cost unitary is implementable with $N$ gates on $N$ layers (App.~\ref{sec:app-clops} provides a faster but more qubit-intensive method). The uncomputation then doubles the gate and layer resources of the QPE part. Altogether, both methods have an asymptotic $\mathcal{O}(N+M)$ layer cost, but IF-QAOA only requires $\mathcal{O}(NM + M^2)$ gates compared to $\mathcal{O}((N+M)^2)$ in the QUBO case.

\paragraph*{Example} Consider a Knapsack instance with $20$ items, $W=200$, $\sum_i w_i = 500$, thus $g^+ = 200$, $g^- = -300$. With $M_\text{QUBO} = 8$, $M_\text{IF} = 9$, QUBO requires $398$ gates and $28$ layers. The QPE part in IF-QAOA requires $20 + 17 = 37$ layers with $180 + 45 = 225$ gates. Consequently, IF-QAOA totals $470$ gates and $94$ layers, clearly showing that IF-QAOA is more resource-intensive in terms of layers. However, the gate count is still comparable.

\subsection{Accelerated Simulation}\label{sec:sim}

Classical simulation of QAOA can be accelerated using precomputation techniques, as explored in Refs.~\cite{lykov2023, juliaqaoa, stein2024a}. These methods are based on the observation that in ordinary state vector simulation, every simulated QAOA layer technically brute-forces the objective function since the values can be read from the phase of the state vector entries. Therefore, brute-forcing the objective function beforehand and reusing it in later iterations is more efficient~\cite{lykov2023, juliaqaoa, stein2024a}. Analogously, we can accelerate the simulation of IF-QAOA by implicitly applying the phase modification in Eq.~\eqref{eq:ifqaoa-phase} without technically simulating the QPE part: namely, we brute-force and cache every element of the cost function $\tilde{f}(x)$ from Eq.~\eqref{eq:if-cost} with the algorithm used in Ref.~\cite{stein2024a}. Then, we multiply $e^{-i\gamma \tilde{f}(x)}$ pointwise to the state vector of the simulator. This is followed by the gate-based application of the $R_X$ mixer gates, similar to~\cite{lykov2023,juliaqaoa,stein2024a}.

With the IF-QAOA, the simulation strategy has the advantage that we are not required to simulate the QPE register qubits, leaving us with a smaller Hilbert space and making larger problem sizes accessible to simulation. The enlarged Hilbert space was precisely why previous methods investigating similar oracle approaches omitted necessary numerical experiments~\cite{kea2023}. 

Additionally, this technique allows the simulation of a penalty-based method that serves as an upper bound in performance to the slack variable-based quadratic penalty method from Eq.~\eqref{eq:slack-penalty}. Namely, we can simulate the cost function
\begin{align}
    f(x) + \lambda (1-\Theta[g(x)])g^2(x),
    \label{eq:virtual-penalty}
\end{align}
subsequently called the \emph{virtual penalty} approach. This cost function is precisely Eq.~\eqref{eq:slack-penalty} if $s(y) = g(x)$ in the feasible part and $s(y) = 0$ in the infeasible part. Here, we again profit from the smaller Hilbert space required for simulation. This allows us to estimate how well the QUBO approach can ideally perform in larger instances, which are impossible to simulate due to the slack variables (and therefore qubits) in place.

\subsection{Approximate Indicator Function}\label{sec:approximate}

Now, we investigate the case when $g(x)$ is not resolvable by the ancillary qubit space, which may be the case if the coefficients are real-valued, i.e., $g(x) \in [g^-, g^+]$, or the number of total qubits is limited. In that case, the Kronecker-Delta identity, used from Eq.~\eqref{eq:ex-qft-applied} to Eq.~\eqref{eq:ex-delta} will not be ideal but approximate. Consequently, there is a marginal probability that the cost function phase is applied even if the constraint is unsatisfied. Nevertheless, we can analytically compute that when we apply the inverse of the QPE and measure the QPE register to be zero, as shown in Fig.~\ref{fig:ifcirc}, only correctly evaluated feasible indicators will get a cost phase
\begin{multline}
    \bra{0}\text{QPE}^\dagger_{U_g} \bar{C}_{z_1}(U_f(\gamma))\text{QPE}_{U_g} \ket{0}\ket{x} \\= \mathcal{P}_M(\gamma f(x), g(x)) \ket{x},\label{eq:projector-defined}
\end{multline}
where $\mathcal{P}_M(\gamma f(x), g(x))$ is a projection factor that depends on the number of ancillary variables $M$. It is visualized in Fig.~\ref{fig:phasefactor} for $M=4$. The projective measurement has a certain probability of failing, i.e., $|\mathcal{P}_M(\gamma f(x),g(x))|^2 \leq 1$, making the operation non-unitary. Since the resolution of QPE grows with increasing ancillary qubits, the probability of a failed projective measurement can be reduced by increasing $M$.

We can analytically express $\mathcal{P}_M$ given by
\begin{multline}
    \mathcal{P}_M(\gamma f(x), g(x)) \\= \frac{e^{-i\gamma f(x)} + 1}{2} + \frac{e^{-i\gamma f(x)} - 1}{2} \vartheta_M(g(x)),\label{eq:projector-main}
\end{multline}
where $\vartheta_M$ is the approximate sign function of $g(x)$, which can be numerically derived from a Fourier series, as detailed in App.~\ref{sec:app-phasemod}. For simulation, we compute $\vartheta_M(g(x))$ beforehand through the brute-forced result of the constraint function $g(x)$, as described in Sec.~\ref{sec:sim}, apply Eq.~\eqref{eq:projector-main} in the optimization.

When simulating the approximate evolution, we have to renormalize the state vector after each application of approximate IF cost layer, denoted by the operator $\mathcal{N}$, yielding the combined layer operation
\begin{align}
\ket{\psi_{i}} = U_M(\beta_{i})\mathcal{N}\mathcal{P}(\gamma_{i} f, g)\ket{\psi_{i-1}},
\end{align}
where $\ket{\psi_i}$ is the quantum state at the end of the $i$-th QAOA iteration. The layer success probability is given by $q_i = \| \mathcal{P}(\gamma_{i} f(x), g(x)) \ket{\psi_{i-1}} \|^2$ and depends on the previous layer, $\gamma_i$, and both functions $f, g$. For many QAOA layers, this leads to an overall success probability that decays like $q_\text{total} = \prod_i q_i$. 


\subsection{Multi-Constraint Setup}

\begin{figure}
    \centering
    \scalebox{0.63}{
    \input{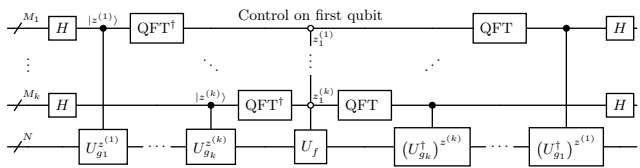}
    }
    \caption{Multi-constraint cost function subroutine. The first qubits of the QPE registers control the application of the cost operator.}
    \label{fig:multi-constrained}
\end{figure}

So far, we have focused only on single constraints.
For the general case with multiple ($k$) inequality constraints, we now propose the following procedure for generating corresponding IF-QAOA circuits, also depicted in Fig.~\ref{fig:multi-constrained}:
\begin{enumerate}
    \item Ensure $f(x) \leq 0 \,\forall x\in \mathcal{D}$. If this is not the case, subtract an upper bound from the cost function. Alternatively, assuring only the optimal solution $f(x^*) \ll 0$ may also suffice.
    \item Compute upper $g^+_i$ and lower $g^-_i$ bounds of all constraint functions. If direct computation of the bounds is impossible, estimate the bounds using, e.g., roof-duality~\cite{hammer1984} or semi-definite programming relaxations~\cite{goemans1995}.
    \item Derive the number of ancillary qubits $M_i$ required for the QPE register of each constraint. If the number of qubits is limited or the constraining function is real-valued, limit the maximal QPE register size to $M_i \leq M_\text{max}$.
    \item Apply the QPEs for each constraint, following Sec.~\ref{sec:ifalgo} and~\ref{sec:approximate}. The QFT part of the QPEs can be executed in parallel.
    \item Apply cost function unitary, multi-controlled by the first qubits of all QPE registers.
    \item Uncompute QPEs and perform the projective measurements if required.
\end{enumerate}
The entire QAOA circuit consists of $N + \sum_i M_i$ qubits and applies the cost function $f(x)\prod_i\Theta[g_i(x)]$. Each QPE requires the application of maximally $\mathrm{deg}(g_i) + 1$ qubit phase gates, and the controlled cost operation consists of maximally $\deg(f) + k$ qubit phase gates, where $\deg(\cdot)$ is the degree of the polynomial.

\paragraph*{Alternative Architectures:}
The multi-controlled gates or the number of qubits may be a bottleneck, therefore we propose two more NISQ-friendly alternatives here: First, instead of relying on the product of the step and cost function, in the no-penalty setting, we can go a step back and introduce constant penalty for every constraint violation, i.e. $f(x) + \sum_i \lambda_i (1-\Theta[g_i(x)])$. This can be implemented by a single phase gate per constraint.
Second, a sequential application of the QPEs can be employed to reduce the qubit count, allowing for the reuse of the QPE register. The no-penalty setting requires storing the indicator information in an additional qubit per constraint, requiring $k-1$ additional qubits. In the penalty setting, however, the sequential application is straightforward.

\paragraph*{Example:}
To illustrate the pipeline, we sketch the procedure for a multi-$k$-Knapsack problem here, defined as
\begin{align*}
    \argmin_x -\sum_{i = 1}^N \sum_{j=1}^k v_i x_{i,j} \quad\text{s.t.}\quad & \sum_i w_{i,j} x_{i,j} \leq W_j\, \forall k \\
    & \sum_k x_{i,k} \leq 1 \,\forall i.
\end{align*}
Since the second type of constraint is a special set-packing constraint that can be enforced through either Grover-Mixers~\cite{bartschi2020} or a special penalty without slack variables~\cite{hess2024}, we only focus on the first Knapsack constraint here.

The Knapsack objective naturally satisfies the first condition in the defined workflow. The second step can be easily evaluated, giving $g_j^- = W_j-\sum_{i}w_{i,j}$ and $g^+_j = W_j$. Then the $M$'s can be evaluated, depending on the preference of $M_\text{max}$, and the circuit can be constructed. Notably, the cost function is separable per Knapsack constraint, allowing us to implement the following effective cost
\begin{align*}
    \tilde{f}(x)=-\sum_k\Theta\left[W_j -\sum_j x_{i,j} w_{i,j}\right]\sum_i v_i x_{i,j} 
\end{align*}
where each Knapsack is taken care of individually.

\section{Numerical Experiments}\label{sec:numerical}
This section presents the numerical simulation experiments conducted to evaluate the performance of IF-QAOA. In our evaluation, we only focus on the single-constraint Knapsack problem instances and rely on ideal statevector simulation with exact (shot-free) evaluation of all quantities, like the expectation value.
All experiments have been conducted on an AMD Ryzen Threadripper PRO 5965WX.

\subsection{Experimental Setup}

Our evaluation compares IF-QAOA against the quadratic penalty approaches as the baseline. We consider both the QUBO method with slack variables (slack penalty) and the virtual penalty from Eq.~\eqref{eq:virtual-penalty} in our experiments. The virtual penalty method serves as an upper performance bound to the QUBO method, helping to access problem instances where the slack qubits are no longer simulatable.

\paragraph{Problem Instances} For the problem sizes $N \in \{6,8,\dots,22\}$, we generate two Knapsack datasets, one with real-valued parameters and one with integer-valued items, both containing 128 instances for each item count. These bias-free instances help to gain insight into the scaling behavior of the algorithms. Starting with the real-valued dataset, we generate the $w_i$ and $v_i \sim \mathcal{U}(0, 1)$. Afterwards, the capacity is sampled from $W \sim \mathcal{U}(0.2,0.8)\sum_i w_i $ to guarantee a binding constraint.

The integer-valued instances are derived from the real-valued ones and use the fixed capacity $10N$. All weights and profits are scaled and rounded to the next integer $w_i \leftarrow \mathrm{round}(10 \times w_iN /W)$, and finally, the capacity is set to $W \leftarrow 10 N$.

Additionally, we normalize the objective function of each approach such that $\max_x \braket{x|H_M|x} - \min_x\braket{x|H_M|x} = 2N = \max_x \tilde{f}(x) - \min_x \tilde{f}(x)$, leading to a more even cost landscape in $\bm{\beta}$ and $\bm{\gamma}$ direction~\cite{montanez-barrera2024}, thereby enhancing optimizer convergence.

\paragraph{Penalty Coefficient}
For the penalty methods from Eq.~\eqref{eq:slack-penalty} and Eq.~\eqref{eq:virtual-penalty}, we must find a penalty coefficient for each problem instance individually. Even though setting $\lambda > \max_i v_i$ is a valid assignment that allows for no infeasible states to have lower energy than the best feasible one~\cite{lucas2014a}, its magnitude may overshadow the optimization objective, potentially leading to suboptimal optimization performance~\cite{ayodele2022a}. To ensure the most optimal performance of the penalty-based approaches in our comparison, we set the penalty such that the lowest energy infeasible state has the same objective value as the second-to-lowest feasible state, i.e.,
\begin{align}
    \min_{x\notin\mathcal{F}} f(x) + \lambda (W-x^Tw)^2 =\min_{x\in \mathcal{F}\setminus\{x^*\}} f(x).
    \label{eq:opt-penalty}
\end{align}
Our proposed parameter setting is empirically validated in App.~\ref{sec:app-penalty}.
As we brute-force prior to QAOA simulation, determining $\lambda$ with our method incurs no additional computational overhead. This approach reduces the magnitude by an average factor of 20 compared to $\max_i v_i$. However, it should be noted that this technique cannot be applied outside the simulation context, which would require reverting to larger, suboptimal penalty factors.

\paragraph{Optimization} The optimization of the QAOA parameters is done with the L-BFGS algorithm~\cite{liu1989}, which has proven to be very efficient in the context of QAOA~\cite{zhou2020a}. The gradients are evaluated exactly based on the adjoint differentiation method, similar to how gradients are computed in the accelerated QAOA framework of Ref.~\cite{stein2024a}. We optimize the parameters for each QAOA layer count $p$ sequentially, starting with $p=1$ and the initial parameter of $\beta_1 = \gamma_1 = 0.1$. After obtaining the optimal parameters, we use linear interpolation for both parameters $\bm{\beta}$ and $\bm{\gamma}$ to the next $p$ as proposed in Ref.~\cite{zhou2020a}. 
Repeating this method, we iteratively optimize for $p \in \{1, 2, 3, 4, 6, 8, 12, 16, 24, 32, 48, 64\}$, with maximally 100 L-BFGS iterations.

Whether IF-QAOA or penalty-based, all methods use the same indicator cost function defined in Eq.~\eqref{eq:if-cost} for training and solution quality evaluation, similar to Ref.~\cite{hess2024}.

\paragraph{Metrics} In line with best practices on benchmarking quantum optimization~\cite{bucher2024}, we will focus on two leading figures of merit to compare the performance of the algorithms. First, the solution quality is measured by computing the \emph{Random-Adjusted} Approximation Ratio (RAAR), defined as follows
\begin{align}
    \text{RAAR} = \frac{\langle\tilde{f}(x)\rangle - \bra{\psi}\tilde{f}(x)\ket{\psi}}{\langle\tilde{f}(x)\rangle -f^*},
\end{align}
where $\langle\tilde f(x)\rangle$ is the random average of the indicator cost and $f^* = \min_{x\in\mathcal{F}} f(x)$ is the optimal solution. Thereby, $\text{RAAR} = 0$ is as good as random sampling, while $\text{RAAR} = 1$ indicates always sampling a perfect solution.

The second metric, Time-to-Solution (TTS), is derived from the probability of sampling the optimal solution $P^* = \left|\braket{x^*|\psi}\right|^2$. We estimate the runtime as the number of shots required until measuring the optimal solution once with 99\% certainty, scaled by the runtime of a single circuit execution. Formally, TTS is defined as follows~\cite{bucher2024,albash2018a}
\begin{align}\label{eq:tts}
    \text{TTS}_p  = \mathbf{L}(p)\left\lceil\frac{\log\,0.01 }{\log(1-P^*)}\right\rceil,
\end{align}
where $\mathbf{L}(p)$ is the number of Circuit Layer Operations (CLOPS) of the depth-$p$ QAOA circuit, which is proportional to the circuit execution time. Please refer to App.~\ref{sec:app-clops} for the precise derivation of $\mathbf{L}(p)$ for both the IF-QAOA and QUBO penalty methods as well as assumptions on gate set and connectivity. The virtual penalty method receives the CLOPS of the slack penalty QAOA. Additionally, we also define the optimal TTS,
\begin{align}
    \text{TTS}^*  = \min_p \text{TTS}_p
\end{align}
as the minimum overall CLOPS required out of the various depth-$p$ experiments.

\subsection{Integer Knapsack Instances}\label{sec:exp-int-ks}
\begin{figure}
\includegraphics[scale=0.8]{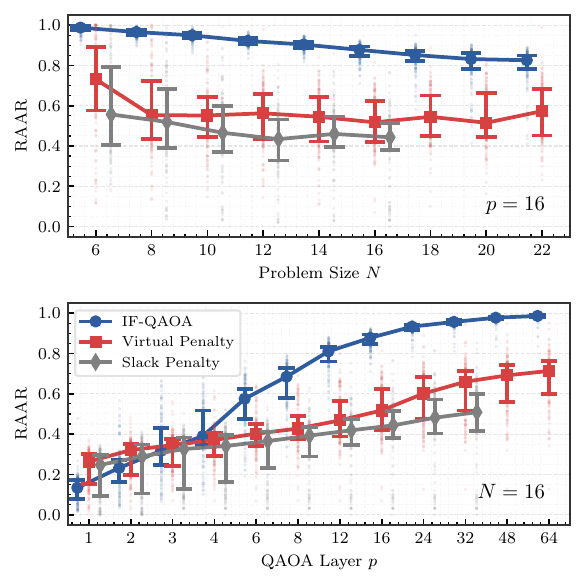}
\caption{The solution quality measured as RAAR displayed for a selection of results. The top plot shows RAAR depending on problem size $N$ for fixed QAOA layer depth $p = 16$, and the bottom plot shows RAAR w.r.t.\ $p$ at $N = 16$. Error bars indicate the 50\% percentile interval.}
\label{fig:int_raar}
\end{figure}

\begin{figure*}
\subfloat[TTS of IF-QAOA and Virtual Penalty\newline w.r.t.\ $p$ for all problem sizes\label{fig:int_tts}]
{\includegraphics[scale=0.73]{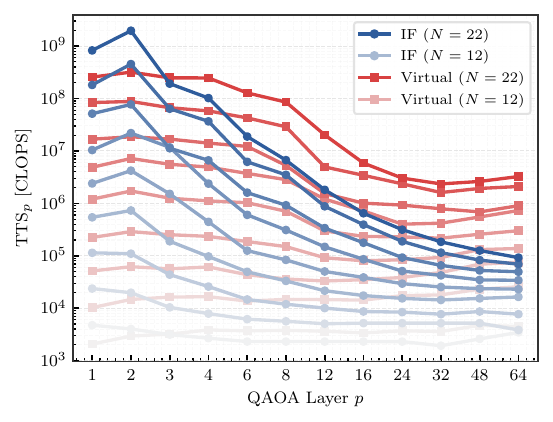}}
\subfloat[TTS$^*$ w.r.t.\ the problem size for all three methods\label{fig:int_tts_opt}]
{\includegraphics[scale=0.73]{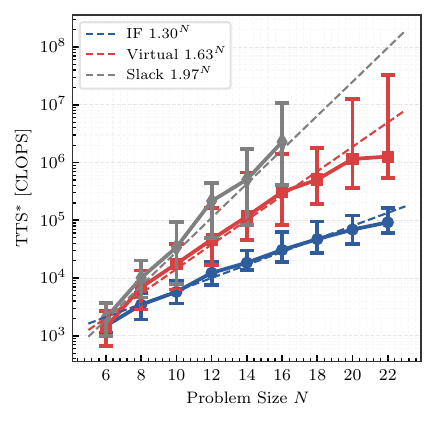}}
\subfloat[TTS$^*$ correlation plot between IF-QAOA and Virtual Penalty\label{fig:int_tts_corr}]
{\includegraphics[scale=0.73]{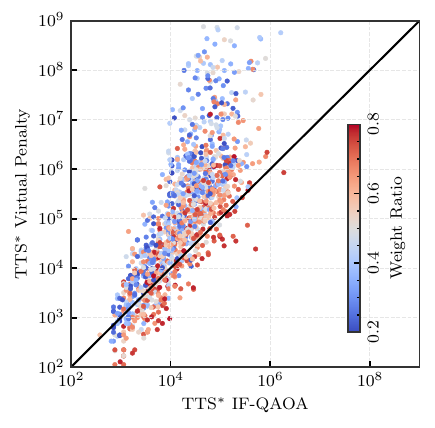}}
\caption{Panel (a) shows the median TTS for all problem instances in various color shades with increasing QAOA layer count $p$. As apparent, IF-QAOA starts with higher TTS than the baseline approach, but starts improving upon it starting with $p \geq 3$. Panel (b) shows the optimal TTS$^*$ depending on the problem size. The dashed line shows the logarithmic regression, with the fitted bases displayed in the legend. Panel (c) shows a scatter plot of all $9 \times 128 = 1152$ problem instances comparing TTS$^*$ of IF-QAOA against the Virtual Penalty. In $82\%$ of the test cases, IF-QAOA arrives at a solution faster. The color shading of the scatter plots indicates the weight ratio $W / \sum_i w_i$ of the Knapsack instance.}
\end{figure*}

We begin the analysis with the fully resolvable integer Knapsack instances results, comparing IF-QAOA against the quadratic penalty approaches.

Fig.~\ref{fig:int_raar} shows slices of the results for both problem sizes $N = 16$ and QAOA layer repetitions $p = 16$. 
It is unambiguously apparent that IF-QAOA produces the best solutions from the compared methods. Focusing on the top panel, showing the results for fixed QAOA repetitions $p=16$ w.r.t.\ problem size, we can observe that IF-QAOA remains at a median $\text{RAAR} > 0.8$ for all problem sizes. In contrast, the penalty-based methods remain in $\text{RAAR} \approx 0.4\text{--}0.6$. Furthermore, the slack penalty method performs slightly worse (but overall similar) to the virtual penalty approach, justifying using the virtual penalty method to estimate how the QUBO method would behave ideally. Besides the median values, it is also evident that IF-QAOA is less dependent on the problem instance itself, which is visible by the smaller Inter-Quartile Ranges (IQRs) marked by the error bars.

When focusing on the RAAR w.r.t.\ the circuit layer depth, it is noteworthy that for low-depth QAOA, the penalty-based methods are outperforming IF-QAOA. We can see improved performance of IF-QAOA starting at depth $p \geq 3$. This could be explained by the heavily penalized infeasible states in the penalty methods, which focus the optimization on finding only feasible states, thereby increasing solution quality fast initially but slowing down once high feasibility has been achieved. IF-QAOA directly emphasizes the better solution candidates, reflecting the improved solution quality at deeper $p$s.
While RAAR continuously improves with increasing $p$ for all methods, IF-QAOA is the only method that approaches perfect solution quality for almost all $N=16$ instances within $p\leq64$.

Fig.~\ref{fig:int_tts} shows the median TTS for IF-QAOA and virtual penalty across all problem sizes and QAOA depths. Similar to the RAAR results in Fig.~\ref{fig:int_raar}, IF-QAOA initially exhibits higher TTS than the virtual penalty counterparts. Yet, starting again with $p \geq 3$, IF-QAOA TTS monotonically decreases and drops below the penalty-based TTS. In contrast, the virtual penalty method shows TTS reduction only for $N > 10$, but this improvement reverses after $p \approx 16$, with the exact threshold depending on the problem size. This reversal means the increased $P^*$ fails to compensate for the longer circuit depth, leading to overall worse TTS runtime.

The optimal TTS$^*$ concerning problem size $N$ is displayed for all approaches in Fig.~\ref{fig:int_tts_opt}, demonstrating the advantageous performance of IF-QAOA. When fitting an exponential to the TTS$^*$ data, IF-QAOA exhibits the most favorable scaling with a base of 1.30 compared to 1.63 of the virtual penalty. The statistical reliability of these fits is supported by the significantly lower variability in IF-QAOA results, which span only half an order of magnitude, versus nearly two orders of magnitude for the virtual penalty method. Interestingly, comparing the fit to the data suggests that both the virtual penalty and IF-QAOA may follow sub-exponential scaling. Further analysis with power-law models reveals that $\text{TTS}^* \propto N^{3.3}$ for IF-QAOA versus $\text{TTS}^* \propto N^{6.1}$ for the virtual penalty.  However, additional experiments on larger problem instances and more carefully selected datasets are necessary to establish confident sub-exponential scaling claims.


\begin{figure}
    \centering
    \includegraphics[width=0.9\linewidth]{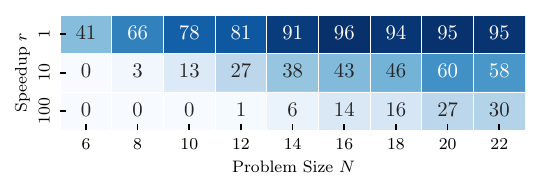}%
    \caption{Share of instances (in \%) that IF-QAOA solves with speedup $r$ compared to the virtual penalty approach, i.e., where $r \mathrm{TTS}^*_\text{IF-QAOA} < \mathrm{TTS}^*_\text{Virtual}$}
    \label{fig:tts_speedup}
\end{figure}

Next, Fig.~\ref{fig:int_tts_corr}, shows a scatterplot of the TTS$^*$ from IF-QAOA against the virtual penalty method, with every dot located above the diagonal line meaning IF-QAOA solves the particular instance faster (TTS$^*$ is lower). Overall, this is the case for 82\% of the instances. The coloring of the data points refers to the weight ratio of that specific instance, defined as $W/\sum_i w_i$. A tendency for cases with a lower weight ratio to be solved faster with IF-QAOA is qualitatively visible. This can be explained by the square of the penalty violation, i.e., $(W - \sum_iw_i)^2$, which can grow much higher for small weight ratio instances. This significantly skews the energy spectrum of the objective function, meaning that finding the optimal solution is less important than finding solutions that only minimize this penalty. IF-QAOA does not change the energy spectrum, making it easier for QAOA to amplify the optimal solution amplitude (which also explains the reduced variance between problem instances). The argument for the energy spectrum can also be made to explain the difference between the slack penalty and virtual penalty approaches. While the virtual penalty only applies correct penalties, the slack penalty introduces additional undesirable energy states. These problematic states arise when the slack variable assignment is inconsistent with the computational register assignment, creating considerable energy contributions that distort the optimization landscape even more.

Fig.~\ref{fig:tts_speedup} gives a more detailed overview of how many instances are solved faster with IF-QAOA compared to the virtual penalty method. Here, we define a speedup factor $r$ and count all instances where $r \mathrm{TTS}^*_\text{IF-QAOA} < \mathrm{TTS}^*_\text{Virtual}$. The results demonstrate that IF-QAOA outperforms the virtual penalty approach for nearly all ($>90\%$) instances with $N \geq 14$. Remarkably, approximately one-third of these larger instances exhibit speedups exceeding two orders of magnitude ($100\times$), highlighting the better scaling behavior of IF-QAOA.

\subsection{Comparison with Quantum Tree Generator Mixer}

\begin{table}
    \centering
    \footnotesize
    \begin{tabular}{l|rrrrrrrrrrrrrrrrrr}
         $N$ & 5 & 6 & 7 & 8 & 9 & 10 & 11 & 12 & 13 & 14 & 15 & 16 & 17 & 18 & 19 & 20 & 21 & 22  \\
         \toprule
         $p$ & 15 & 20 & 22 & 28 & 28 & 40 & 41 & 45 & 52 & 46 & 50 & 61 & 57 & 70 & 64 & 64 & 69 & 69
    \end{tabular}
    \caption{Number of IF-QAOA repetitions until CLOPS of QTG-QAOA at $p=1$ is matched.}
    \label{tab:qtg}
\end{table}

\begin{figure}
    \centering
    \includegraphics[scale=0.8]{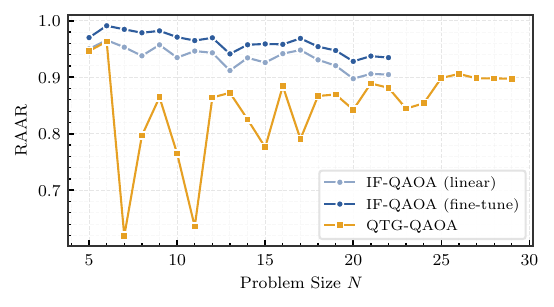}
    \caption{RAAR of IF-QAOA (for $p$, see Tab.~\ref{tab:qtg}) compared to QTG-QAOA at $p=1$. IF-QAOA is separated between two results: linear refers to optimizing just the rates of the linear schedule. Fine-tune refers to running 100 L-BFGS iterations on all parameters, starting with the optimized linear schedule. Results for QTG-QAOA and problem instances are taken from Ref.~\cite{christiansen2024}.}
    \label{fig:qtg-comparison}
\end{figure}

To benchmark IF-QAOA against leading constrained preserving architectures, we selected the QTG-based AAM-QAOA (QTG-QAOA) as the most promising architecture for Knapsack problems~\cite{christiansen2024}. The authors showed a clear performance advantage of QTG-QAOA, outperforming Ref.~\cite{dam2022}. We build our comparison on their results, featuring hard Knapsack instances generated by Ref.~\cite{jooken2022} and depth $p=1$ QTG-QAOA only.

Solving the same Knapsack instances~\cite{christiansen2024}, we set the number of IF-QAOA layers such that the total IF-QAOA circuit requires just as many CLOPS as $p=1$ QTG-QAOA, i.e.,
\begin{align}
    \mathbf{L}_\text{IF}(p) \leq \mathbf{L}_\text{QTG}(1).
\end{align}
This way, we can ensure that better solution quality means better performance.
The resulting $p$-values are summarized in Table~\ref{tab:qtg}, showing an increasing number of repetitions possible for IF-QAOA due to the more streamlined circuit architecture.

Since we instantly aim to optimize for large $p$ values (Table~\ref{tab:qtg}), we skip the previous interpolation scheme. Instead, we first optimize a linear schedule similar to Ref.~\cite{sack2021}
\begin{align}
    \gamma_i = \Delta_\gamma \frac{i+1/2}{p} \qquad
    \beta_i = \Delta_\beta \frac{p-i-1/2}{p},
\end{align}
where $\Delta_{\gamma,\beta}$ are the two parameters to be tuned. Fig.~\ref{fig:qtg-comparison} shows the RAAR after 100 L-BFGS iterations, which is clearly above the RAAR found by the QTG approach. When fine-tuning each of the $\beta_i,\gamma_i$ starting with the optimized linear schedule, the RAAR can be improved even more. Yet, since we only simulate until $N=22$, we cannot predict whether this pattern continues or QTG provides advantageous results for larger instances. \citeauthor{christiansen2024}~\cite{christiansen2024} were able to simulate the larger problems since QTG restricts state vector simulation to the feasible subspace. These results indicate that simpler circuit architecture in the context of QAOA can lead to better results overall since more repetitions of a single QAOA layer can be employed to find the optimal solution directly. However, there is a break-even to be expected since the QTG-based approach only searches in the space of feasible assignments, while the IF-QAOA needs to find feasible results through optimization. 
\begin{figure*}[ht]
\subfloat[Success probability $q_\text{total}$ over all instances depending on $p$ and $M$\label{fig:qpe_psucc}]
{\includegraphics[scale=0.73]{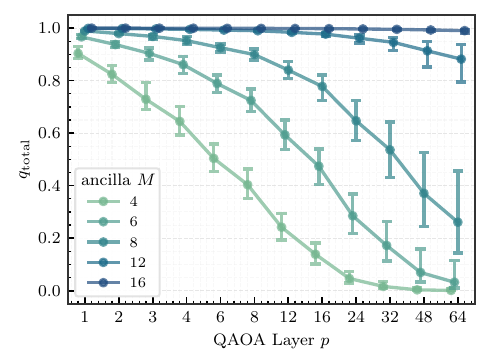}}
\subfloat[TTS at problem size $N = 16$ depending\newline on $p$ and $M$\label{fig:qpe_tts}]
{\includegraphics[scale=0.73]{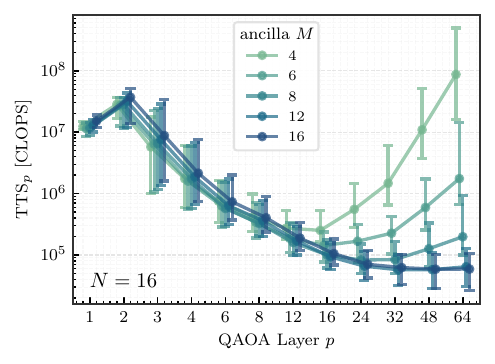}}
\subfloat[TTS$^*$ scaling w.r.t.\ problem size\label{fig:qpe_tts_scaling}]
{\includegraphics[scale=0.73]{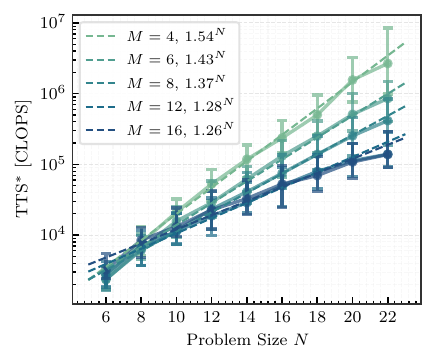}}
\caption{The success probability $q_\text{total}$ for different $M$ is depicted in panel (a). Panel (b) shows the TTS w.r.t.\ the QAOA repetitions for problem size $N=16$. Clearly, fewer ancillary variables lead to a sudden increase in TTS after $p \approx 16$ due to vanishing success probability. TTS is relatively similar for shorter QAOA circuits. TTS$^*$ scaling is shown in panel (c), with more ancillary variables leading to more favorable scaling. However, we also see that the scaling saturates and is very comparable for $M \geq 12$.}
\end{figure*}

\subsection{Approximated IF Results}

As the next and final analysis of the numerical results, we focus on the non-integer Knapsack instances and approximative IF application.
Since the approximate IF is not sharp like the ideal step function, see Fig.~\ref{fig:phasefactor}, we introduce a small offset of the constraining function, shifting the transition region more to the center, not occurring between $g(x) \in [-1, 0]$, but around $g(x)\approx0$. Namely, we implement $\Theta[g(x) - \epsilon]$, where $\epsilon \in [0,1]$. Running a linear search on a subset of the instances revealed that $\epsilon=0.5$ empirically works best. The effect of the offset decreases with increasing resolution in the QPE register.

Since the projective measurement only has a certain probability of succeeding per QAOA repetition, here given as $q_i$, we expect the overall success probability $q_\text{total} = \prod_i q_i$ to deteriorate with an increasing number of repetitions and approximation degree. This can precisely be observed in Fig.~\ref{fig:qpe_psucc}, showing $q_\text{total}$ w.r.t.\ $p$. The higher the number of ancillary variables is in the QPE register for IF computation, the lower the approximation degree of the sign function becomes and, therefore, also the lower any probability of failing.

Since we now have failure probability during circuit execution, TTS from Eq.~\eqref{eq:tts} requires redefinition. Whenever the projective measurement fails, we can instantly start over, leaving us with an expected circuit layer depth of
\begin{align}
    \langle\mathbf{L}\rangle(q, p) &= \mathbf{L}_\text{init} + \mathbf{L}_\text{layer} (1 + q_1 + q_1 q_2  + q_1 q_2 q_3 + \cdots)\nonumber\\
    &= \mathbf{L}_\text{init} + \mathbf{L}_\text{layer} \left(1 +\sum_{i=1}^{p-1} \prod_{j=1}^i q_i\right).
\end{align}
Together with the combined probability of successful indicator applications and finding the optimal solution $P^* q_\text{total}$, we can define the TTS for the approximative IF-QAOA as follows
\begin{align}
    \text{TTS}_p = \langle \mathbf{L} \rangle(q,p) \left\lceil\frac{\log 0.01}{\log(1 - P^*q_\text{total})}\right\rceil,
\end{align}
which is shown in Fig.~\ref{fig:qpe_tts} for problem size $N= 16$. Here, we can see that increased probability decay throughout the computation impacts TTS significantly with small QPE registers. For instance $M=4$ exhibits slower TTS for $p = 64$ compared to $p = 1$. The detrimental effect of projective measurements on TTS diminishes steadily as $M$ increases. At $M \geq 12$, the performance impairment becomes negligible. Consequently, as $M$ increases, TTS$^*$ becomes smaller and the optimal depth $p$ shifts to the right.

The same can be observed when considering the scaling behavior of TTS$^*$, as shown in Fig.~\ref{fig:qpe_tts_scaling}. Higher resolution in the QPE register leads to more favorable scaling. The bases for the exponential fit decrease with increasing $M$ from 1.54 at $M=4$ to 1.26 at $M=16$. Notably, even at $M=4$, the TTS$^*$ scaling outperforms the penalty approaches discussed in Sec.~\ref{sec:exp-int-ks}. Also, scaling improves upon the exact integer cases, which can be explained by the non-growing complexity of the oracle circuit due to fixed $M$. Furthermore, a sub-exponential behavior can also be observed here. A power-law fit reveals $\text{TTS}^* \propto N^{2.9}$ at $M=16$.

At $M=12$ and $M=16$, the exponential bases are very similar (1.28 and 1.26), affirming our prior statement that a resolution of $M\approx12$ suffices for our benchmark instances. It is likely that when increasing $N$ to sizes beyond our benchmark instances $M$ has to be further increased. To solve larger instance, $p$ must grow beyond the tested depths, which subsequently causes the $q_\text{total}$ to decay, requiring more resolution.

\section{Discussion \& Conclusion}\label{sec:discussion}
We have provided a formalized description of the IF-QAOA for general inequality constraint encoding of COPs using a low-complexity oracle circuit consisting of two QPEs. This circuit allows us to compute constraint satisfaction onto a single indicator qubit, which can subsequently be used to apply the cost function in a controlled fashion, not requiring penalty coefficients. We provided accelerated simulation techniques for both the exact and the approximate cases, where the image of the constraining function is not fully resolvable in the QPE register. Finally, we evaluated the performance of IF-QAOA compared to the default QUBO method, i.e., adding a slack variable and a quadratic penalty to the objective, using randomly generated Knapsack instances.

We found that IF-QAOA produces more favorable results with a better approximation ratio and faster optimal TTS. Overall, in 82\% of the tested instances, IF-QAOA arrived at the solution faster, despite slightly higher circuit layer complexity than the QUBO implementation due to oracle construction. Also, we found that the empirical scaling behavior of IF-QAOA has a lower exponential base (1.30) than the quadratic virtual penalty method (1.61). Investigating the instance properties, it was apparent that the weight ratio significantly impacted deciding which method solves a problem faster. Small weight ratios result in large penalties that significantly distort the energy spectrum. This distortion causes the quadratic penalty method to prioritize constraint satisfaction over finding the optimal solution within the feasible space. In contrast, IF-QAOA preserves the original energy spectrum, focusing on solution quality.

Additionally, we compared IF-QAOA against the state-of-the-art AAM-QAOA based on QTG for Knapsack~\cite{christiansen2024, wilkening2024}. Since the QTG-QAOA overhead allows for multiple IF-QAOA iterations to match circuit complexity, we found that IF-QAOA produces better results regarding RAAR when using a similar number of CLOPS. While our comparison with QTG-QAOA showed promising results for IF-QAOA at smaller problem sizes, we acknowledge the limited scope of this comparison. The generalizability of IF-QAOA's performance advantage remains an open question for larger problem instances, where QTG's restricted search space approach may ultimately prove more efficient. Future work should include more extensive benchmarking against ($p > 1$) QTG-QAOA across a broader range of problem sizes to establish more precise performance boundaries between these approaches. Nevertheless, it is worth noting that IF-QAOA is generally applicable compared to the Knapsack-focused QTG.

Concluding the results, IF-QAOA in the approximate regimes proved that even a small number of ancillary qubits suffice to find solutions even though each layer has some failure probability. For $M \geq 12$, the resolution of the approximate sign function was satisfactory for the considered benchmarking instances. Increasing $M$ did not improve performance anymore.

We can confidently conclude from our results that IF-QAOA provides a practical, low-complexity framework with promising performance results for integrating inequality constraints into QAOA and solving general forms of COPs. However, this claim sustains itself solely on simulations of the Knapsack problem, the simplest constrained optimization problem. An evaluation on a broader range of issues is required to reinforce this statement. The main limitation of IF-QAOA is that it is an entirely cost-function-based method, meaning that QAOA optimization is necessary to find feasible solutions. This can become detrimental when the ratio of feasible solutions compared to the search space diminishes. Another technical drawback is that the entire image of the constraining function needs to be resolved in the non-approximative case, compared to only the feasible range of the QUBO method, leading to a potential overhead in qubit resources.

For future work, IF-QAOA has to be tested in a multi-constraint setting with the Multi-Knapsack formulation already given in Sec.~\ref{sec:methods}. Furthermore, IF-QAOA's flexibility can be taken advantage of by combining it with different constraint-preserving techniques, e.g., $XY$-Mixers~\cite{wang2020} or Grover Mixers~\cite{bartschi2020}. For instance, the second constraint in the Multi-Knapsack problem can be enforced using Grover-Mixers.
Experiments on hardware need to prove that the higher ideal optimization quality can remain in the presence of noise. The slightly higher circuit layer complexity can be obtrusive in this case, but competitive results can still be expected since the gate count scales similarly to QUBO. 
Furthermore, different circuits can entirely replace the QPE when implementing the adder in the IF circuit. For instance, one could imagine Quantum Signal Processing (QSP)~\cite{low2017} to construct the step function. Additionally, Quantum Singular Value Transform (QSVT)~\cite{gilyen2019} constructions could implement more complex or specialized constraining functions in the ancillary register, e.g., constraints based on the solution of linear systems of equations, similar to simulation-based optimization~\cite{stein2024b}. 

\section*{Code \& Data Availability}
The code and data for all experiments is available under \href{https://github.com/dbucher97/fast-qaoa-c}{github.com/dbucher97/fast-qaoa-c}.

\begin{acknowledgments}
This work was supported by the German Federal Min-
istry of Education and Research under the funding program
“Förderprogramm Quantentechnologien – von den Grundlagen
zum Markt” (funding program quantum technologies — from
basic research to market), project QuCUN, 13N16199.    
J.S. acknowledges support from the German Federal Ministry for Economic Affairs and Climate Action through the funding program "Quantum Computing – Applications for the industry" based on the allowance "Development of digital technologies" (contract number: 01MQ22008A).
\end{acknowledgments}

\appendix

\section{Derivation of Projection Factor in Approximate IF-QAOA}\label{sec:app-phasemod}

To derive $\mathcal{P}_M$ and $\mathcal{\vartheta_M}$ in Eq.~\eqref{eq:projector-main}, we begin with the expression in Eq.~\eqref{eq:ex-qft-applied}, where the QPE has already been applied:
\begin{align}\label{eq:app-begin}
    \frac{1}{2^M} \sum_{z = 0} \sum_{k = 0} e^{-\frac{2 \pi i}{2^M} (z - g(x))k}  \ket{x}\ket{z} = \ket{\psi(x)}.
\end{align}
Then we apply the controlled evolution of the cost Hamiltonian, which splits the term into two sections: One, where $z_1 = 0$, the constraint is satisfied, and the cost function is applied, and where $z_1 = 1$, the constraint is violated, and no phase is altered 
\begin{align}
    &\ket{\psi'(x)} = \bar{C}_{z_1}(U_f) \ket{\psi(x)}  \\
    &=\frac{1}{2^M} \sum_{z = 0} \sum_{k = 0} e^{-\frac{2 \pi i}{2^M} (z - g(x))k} e^{-i \gamma f(x) (1- z_1)} \ket{x} \ket{z}.\nonumber
\end{align}

As defined in Eq.~\eqref{eq:projector-defined}, the projection factor $\mathcal{P}_M$ it is given by the overlap of $\ket{\psi}$ and $\ket{\psi'}$:
\begin{align}
    &\mathcal{P}_M(\gamma f(x) , g(x)) = \braket{\psi(x)|\psi'(x)} \\
    &= \frac{1}{2^{2M}} \sum_{k,k'} e^{\frac{2 \pi i}{2^M} g(x) (k - k')} \sum_{z} e^{-\frac{2 \pi i}{2^M} z (k - k')} e^{-i \gamma f(x)(1-z_1) }\nonumber\\
    &= \frac{1}{2^M} \sum_{k,k'} e^{\frac{2 \pi i}{2^M} g(x) (k - k')} \phi(k - k'),\label{eq:app-pr-phi}
\end{align}
with helper function $\phi(k)$ defined for $-2^M < k < 2^M$. This function can be further simplified, leading us to
\begin{align}
    \phi(k) &= \frac{1}{2^M} \sum_{z = 0} e^{-i \gamma f(x) (1 - z_1)} e^{-\frac{2 \pi i}{2^M} z k} \\
    &= \frac{1}{2^M} \sum_{z = 0}^{2^{M - 1} - 1} \left(e^{-i \gamma f(x)}  + e^{\frac{2\pi i}{2} k} \right) \left(e^{-\frac{2\pi i}{2^M} k}\right)^z \label{eq:app-pr-1}\\
    &= \frac{e^{-i \gamma f(x)} + (-1)^k}{2^M} \frac{1 - e^{-\pi i k}}{1 - e^{-\frac{2 \pi i} {2^M} k}}\label{eq:app-pr-2} \\
    &= \begin{cases}
        \frac{1}{2} \left(e^{-i\gamma f(x)} + 1\right )  &\text{if }k = 0 \\
        \frac{1}{2^M} \frac{2}{1 - e^{-\frac{2\pi i}{2^M} }} \left(e^{-i\gamma f(x)} - 1\right )  &\text{if $k$ odd} \\
        0  &\text{else},
    \end{cases}
\end{align}
using the geometric sum formula between Eq.~\eqref{eq:app-pr-1} and Eq.~\eqref{eq:app-pr-2}. Since the summand in Eq.~\eqref{eq:app-pr-phi} only depends on $k - k'$ we can substitute $k - k' \rightarrow k$ together with a weighting factor $(2^M - |k|)$, leading to
\begin{align}
    &\mathcal{P}_M(\gamma f(x), g(x)) = \phi(0) -  \sum_{k = 0}^{2^{M - 1} - 1} \left(1 - \frac{2 k + 1}{2^M}\right) \nonumber\\&\left[ e^{\frac{2\pi i}{2^M} g(x) (2 k + 1)}  \phi(2 k + 1) + e^{-\frac{2\pi i}{2^M} g(x) (2k+1)}  \phi(-2 k - 1) \right]\nonumber\\
    &= \frac{e^{- i \gamma f(x)} + 1}{2} + \frac{e^{-i \gamma f(x)} - 1}{2} \vartheta_M(g(x)),
\end{align}
with
\begin{multline}
    \vartheta_M(g(x)) = \\\frac{2}{2^{M-2}} \mathrm{Re} \sum_{k = 0}^{2^{M-1} - 1} \left(1 - \frac{2 k + 1}{2^M}\right) \frac{e^{\frac{2\pi i}{2^M} g(x)(2k+1)}}{1 - e^{-\frac{2 \pi i}{2^M} (2k +1) }}.\label{eq:app-final}
\end{multline}
The approximate sign function $\vartheta_M(\ell) = \mathrm{sign}(\ell)$ if $\ell \in \{-2^{M-1},\dots ,2^{M-1} -1\}$. Since any $g(x) \in [g^-, g^+]$ can be expressed as $\ell + \delta$ with $\delta \in [0,1)$, we can identify \eqref{eq:app-final} with an inverse Fourier transform to compute unit spaced values offset by $\delta$ simultaneously, i.e. 
\begin{align}
    \vartheta_M(\ell + \delta) = 2 \mathrm{Re}\,\mathcal{F}^{-1}_{k\rightarrow \ell}\left[(1 - k/2^M)\tilde{\phi}(k)e^{\frac{2\pi i}{2^M} \delta k}\right],
\end{align}
with $\tilde{\phi}(0) = 0$ and $\tilde{\phi}(k) = \phi(k)$, otherwise.
This allows fast and accurate numeric derivation of multiple $\vartheta_\ell[g(x)]$ terms since we can pre-compute various offset factors with Fast Fourier Transform (FFT) beforehand and interpolate them afterwards. From the $\mathcal{P}$ plot in Fig.~\ref{fig:phasefactor}, it is apparent that the oscillations of the $\vartheta$ are of length 1, therefore, $2^m = 4$ super-samples ($\delta \in \{0, 1/4, 1/2, 3/4\}$) already allow for accurate interpolation of any $g(x)$. With higher $m$, the numerical error decreases at the cost of more FFT calls. In total, the computational complexity for evaluating all support points is given by $\mathcal{O}(M2^{M+m})$ with $m$ being generally small. Throughout this work, we use a super-sampling factor of $m = 3$.

Note that the geometric sum formula can also already be computed in Eq.~\eqref{eq:app-begin}, which leads to a more manageable expression right away. However, this leads to an expression of $\mathcal{P}$ that does not allow for the Fourier transform trick, giving us inferior runtime complexity of $\mathcal{O}(2^{2M + m})$ when interpolating, or $\mathcal{O}(2^{M + N})$, when evaluating $\vartheta(g(x))$ directly from the brute-forced $g(x)$ values.

%
%


\section{Empirical Penalty Factor Verfication}\label{sec:app-penalty}

\begin{figure}
    \centering
    \includegraphics[width=\linewidth]{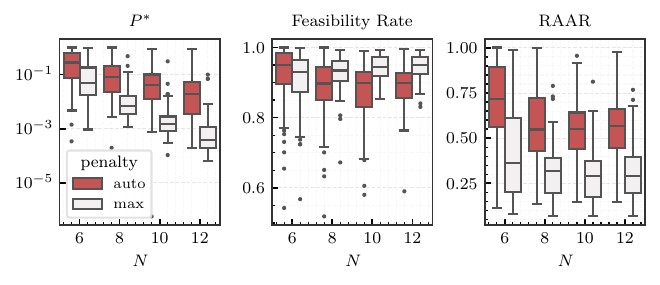}%
    \caption{Probability of sampling the optimal solution, feasibility rate $\sum_{x\in\mathcal{F}} \left|\braket{x|\psi}\right|^2$, and RAAR at $p=16$ for different problem sizes w.r.t.\ penalty settings, where auto refers to Eq.~\eqref{eq:opt-penalty}.}
    \label{fig:app-feas-rate}
\end{figure}

Fig.~\ref{fig:app-feas-rate} shows the performance of the virtual penalty approach with the penalty setting as explained in Eq.~\eqref{eq:opt-penalty} (auto) and penalty set to $\lambda = \max_i v_i$ (max).

The probability of sampling the optimal solution $P^*$ (left plot), which later determines TTS, is significantly higher with the auto penalty method, clearly showing that our approach, explained in Sec.~\ref{sec:numerical}, is a valid choice for penalty determination.

However, as the penalty is set to a smaller value than $\max_i v_i$, infeasible states may have a less objective value than some feasible states, which means that a smaller fraction of feasible states will be measured overall. This is verified by the data in the center plot, where the max method exhibits a higher feasibility rate.

Conversely, due to the less overshadowed objective function, the auto method emphasizes on better solutions, resulting in overall better solution quality in terms of RAAR (right plot).

\section{Resource Estimation}\label{sec:app-clops}
    
To reliably compute the circuit resources in terms of Circuit Layer Operations (CLOPS), we first need to define what gates are considered a single circuit layer operation. Like Ref.~\cite{wilkening2024}, we define single-controlled Pauli rotations and CNOT gates as single-layer operations. Likewise, we also assume all-to-all connectivity, meaning that we can place gates between arbitrary qubits as long as one of the qubits is not yet used in that layer. The general QAOA circuit resources can be expressed as follows:
\begin{align}
    \mathbf{L}_\text{QAOA}(p) = \mathbf{L}_\text{init} + p(\mathbf{L}_\text{cost} + \mathbf{L}_\text{mixer})
\end{align}
where $\mathbf{L}_\text{mixer} = 1$, $\mathbf{L}_\text{init} = 1$ in both the QUBO and IF case.

That leaves us with finding the resources to implement the cost layer. Due to the fully connectedness of the QUBO with slack variable from Eq.~\eqref{eq:slack-penalty}, 
\begin{align}
    \mathbf{L}_\text{cost}^\text{QUBO} = N+M\,\, (-1),
\end{align}
where 1 is subtracted if $N + M$ is even given by the edge colorability of a complete graph~\cite{lint2006}.

In IF-QAOA, the cost layer is more complex to evaluate
\begin{align}
    \mathbf{L}^\text{IF}_\text{cost} = 2 \mathbf{L}_\text{add} + 2\mathbf{L}_\text{QFT}  + \mathbf{L}_\text{penalty},
\end{align}
where $\mathbf{L}_\text{add}$ is the CLOPS for adding the weight contributions in the phase of the QPE register. The QFT resource is given by $\mathbf{L}_\text{QFT} = 2M - 1$ through parallel execution~\cite{wilkening2024}.

The addition gates can be seen as an edge coloring problem of the graph connecting all compute qubits with all QPE qubits. The chromatic number subsequently gives the number of parallel rotation gate executions. As the connection graph is entirely bipartite, K\H{o}nigs theorem tells that the chromatic number is the maximum degree in the graph~\cite{biggs1998}, which yields in our case
\begin{align}
    \mathbf{L}_\text{add} = \max\{N, M\}.
\end{align}

The cost part is given by $N$ sequential controlled rotations applied on the compute register qubits
\begin{align}
    \mathbf{L}_\text{cost} = N.
\end{align}
Alternatively, the controlled rotation can be accelerated by copying the state of the indicator qubit onto $N-1$ additional qubits, allowing for
\begin{align}
    \mathbf{L}_\text{cost} = 2\ceillog{N} + 1,
\end{align}
as proposed in Ref.~\cite{wilkening2024}.

In total, the cost layer of the IF-QAOA has the following defined resources
\begin{align}
    \mathbf{L}_\text{cost}^\text{IF} = 2 \max\{N, M\} + 4M + 2\ceillog{N} - 1.
\end{align}

\bibliography{main}

\end{document}